\documentclass[aps,reprint,amsmath,amssymb,superscriptaddress,floatfix,twocolumn]{revtex4-1}
\usepackage{amsmath}
\usepackage{changepage}
\usepackage{epsfig}
\usepackage{epstopdf}
\usepackage[colorlinks,citecolor=blue,linkcolor=blue]{hyperref}
\usepackage{lipsum}
\usepackage{booktabs}
\usepackage{multirow}
\usepackage{braket} 
\usepackage{multirow}
\usepackage{makecell}
\usepackage{amsmath}
\usepackage{mathtools}

\begin{document}
\title{Interaction between giant atoms in a one-dimensional topological waveguide}
\author{Da-Wei Wang}
\affiliation{School of Physics, Dalian University of Technology, Dalian 116024,People's Republic of China}
\author{Chengsong Zhao}
\affiliation{School of Physics, Dalian University of Technology, Dalian 116024,People's Republic of China}
\author{Junya Yang}
\affiliation{School of Physics, Dalian University of Technology, Dalian 116024,People's Republic of China}
\author{Ye-Ting Yan}
\affiliation{School of Physics, Dalian University of Technology, Dalian 116024,People's Republic of China}
\author{Zhihai-Wang}
\affiliation{Northeast Normal University, Changchun 130024,People's Republic of China}
\author{Ling Zhou*}
\affiliation{School of Physics, Dalian University of Technology, Dalian 116024,People's Republic of China}
\thanks{zhlhxn@dlut.edu.cn}

\begin{abstract}

  In this paper, we consider giant atoms coupled to a one-dimensional topological waveguide reservoir. We studied the following two cases.
  In the bandgap regime, where the giant-atom frequency lies outside the band, we study the generation and distribution of giant atom-photon bound states and the difference between the topological waveguide in topological and trivial phases.   When the strengths of the giant atoms coupled to the two sub-lattice points are equal, the photons distribution is symmetrical and the  chiral photon distribution is exhibited when the coupling is different. The coherent interactions between giant atoms are induced by virtual photons, or can be understood as an overlap of photon bound-state wave functions, and decay exponentially with increasing distance between the giant atoms. We also find that the coherent interactions induced by the topological phase are larger than those induced by the trivial phase for the same bandgap width.
  In the band regime, the giant-atom frequency lies in the band, under the Born-Markov approximation, we obtained effective coherence and correlated dissipative interactions  between  the giant atoms mediated by topological waveguide reservoirs, which depend on the giant-atom coupling nodes.  
  We analyze the effect of the form of the giant-atom coupling point on the decay, and on the associated dissipation.  The results show that we can design the coupling form as well as the frequency of the giant atoms to achieve zero decay and correlation dissipation and non-zero coherent interactions.  Finally we used this scheme to realize the excitation transfer of giant atoms. Our work will promote the study of topological matter coupled to giant atoms.

\end{abstract}

\maketitle
\section{Introduction}

With the rapid development of micro- and nano-manufacture technology, waveguide quantum electrodynamics (waveguide QED) involving the coupling of atoms and one-dimensional propagation fields has become an important physical platform for realizing quantum information processing  \cite{Brehm2021,Zanner2022,goban2014atom,PhysRevLett.115.063601,RevModPhys.89.021001,PhysRevLett.111.053601,doi:10.1126/science.ade7651,doi:10.1126/science.1244324} and quantum simulation  \cite{Douglas2015, RevModPhys.86.153,PhysRevX.11.011015}. When the transition frequencies of atoms lie in the photonic band gap, photons are localized around them, which form bound states that can mediate coherent interactions between atoms \cite{Liu2017, Li2019,PhysRevLett.123.233602,PhysRevLett.64.2418,PhysRevX.6.021027}. 
This yields many interesting phenomena and applications such as the generation of long-range entanglement \cite{PhysRevLett.110.080502,PhysRevLett.106.020501}, photon transport \cite{PhysRevX.5.041036,PhysRevA.82.063816}, unconventional quantum optical \cite{Bello2019, doi:10.1126/science.ade9324,PhysRevLett.126.203601,PhysRevA.104.053522,PhysRevResearch.4.023077,PhysRevResearch.3.013025} and the simulation of topological states \cite{Ringel_2014,RevModPhys.95.015002,PhysRevB.105.094422,PhysRevA.82.063816}.

Waveguide quantum electrodynamics (Waveguide QED), which involves the coupling of atoms to a one-dimensional propagating field, has become an important physical platform for realizing quantum information processing and quantum simulation with the rapid development of micro- and nano-fabrication technologies. When waveguides are designed with finite bandwidths and non-linear dispersion relations, the physics of light-matter interactions in one dimension becomes  complicated and interesting such as coupled cavity arrays described by the tight-binding model \cite{PhysRevLett.124.213601,PhysRevA.93.033833,PhysRevLett.111.103604,PhysRevLett.101.100501,PhysRevA.83.043823} and one-dimensional topological photonic baths described by the Su-Schrieffer-Heeger (SSH) model \cite{PhysRevLett.42.1698,RevModPhys.88.021004}.
In  recent years, topological photonics have received a great deal of interest and attention in quantum physics because of their many interesting features, including robustness to local decoherence processes and potential applications in quantum information \cite{RevModPhys.91.015006,RevModPhys.82.3045,Shalaev2019,Wray2010}.  Topological waveguide described by the SSH model  has been realized in superconducting circuit\cite{PhysRevX.11.011015} and photonic waveguide \cite{PhysRevLett.129.173601}. Meanwhile,  atoms coupled with topological waveguide present new opportunities to study single-excitation dynamics, topologically protected states, and mediated interactions between atoms  \cite{PhysRevLett.126.063601,PRXQuantum.3.010336,PhysRevX.11.011015,PhysRevA.104.053522,PhysRevLett.124.023603,PhysRevApplied.15.044041,PhysRevResearch.2.012076}.  In particular,  several unconventional quantum optical phenomena were predicted and realized when quantum emitters interact with a topological waveguide,  such as the emergence of chiral bound states and topologically dependent super/sub-radiative states \cite{Bello2019,PhysRevX.11.011015}.

In conventional waveguide-QED systems, atoms are considered  point-like objects during their interaction with the waveguide due to the assumption that the wavelength of the atoms is much smaller than the wavelength of the coupling field, this is known as the dipole approximation \cite{scully_zubairy_1997}. Recently, a new model of giant atoms has been developed in artificial atomic systems, in which superconducting quantum qubits are exploited as giant atoms \cite{PhysRevA.90.013837,PhysRevA.95.053821,Kannan2020}.  
Furthermore, it has recently been reported that the structure of giant atoms can also be implemented in the optical lattice of cold atoms \cite{PhysRevLett.122.203603}, the giant spin ensemble \cite{dwang2022} and synthetic frequency dimension \cite{PhysRevLett.128.223602}.

Compared to  point-like atoms, the giant atoms can be conjoined to the waveguide by multiple coupling points, which leads to many peculiar phenomena due to the interference and time delay effects between the coupling points such as  frequency-dependent decay rates and Lamb shifts \cite{PhysRevLett.120.140404,Andersson2019}, waveguide-mediated decoherence-free interaction \cite{PhysRevA.107.013710,PhysRevA.107.023705,PhysRevResearch.2.043184}, chiral phenomenons \cite{PhysRevA.105.023712,PhysRevLett.126.043602,Jia2023,PhysRevA.103.053701}, and oscillating bound states \cite{PhysRevResearch.2.043014,PhysRevA.107.023716,PhysRevA.102.033706}. Meanwhile, various meaningful theoretical proposals have recently been suggested by considering the giant atoms with time-dependent couplings, and modulated transition frequency \cite{PhysRevResearch.4.023198}.
When the giant atoms couple to the  topological waveguide,  both the giant atoms and the photon behave in exotic ways, for example, the giant atom can act as an effective boundary and induce the chiral zero-energy modes for the topological waveguide under the periodic boundary \cite{PhysRevA.106.033522}.

In this paper, we consider giant atoms coupled to a one-dimensional topological waveguide reservoir. We studied the following two cases.
In the bandgap regime, where the giant-atom frequency lies outside the band, we study the generation and distribution of giant atom-photon bound states and the difference between the topological waveguide in topological and trivial phases.   When the strengths of the giant atoms coupled to the two sub-lattice points are equal, the photons distribution is symmetrical and the  chiral photon distribution is exhibited when the coupling is different. The coherent interactions between giant atoms are induced by virtual photons, or can be understood as an overlap of photon bound-state wave functions, and decay exponentially with increasing distance between the giant atoms. We also find that the coherent interactions induced by the topological phase are larger than those induced by the trivial phase for the same bandgap width.
In the band regime, the giant-atom frequency lies in the band, under the Born-Markov approximation, we obtained effective coherence and correlated dissipative interactions  between  the giant atoms mediated by topological waveguide reservoirs, which depend on the giant-atom coupling nodes.  
We analyze the effect of the form of the giant-atom coupling point on the decay, and on the associated dissipation.  The results show that we can design the coupling form as well as the frequency of the giant atoms to achieve zero decay and correlation dissipation and non-zero coherent interactions.  Finally we used this scheme to realize the excitation transfer of giant atoms.

\section{The System}


\begin{figure}[h]
  \centering
  \includegraphics[width=8cm]{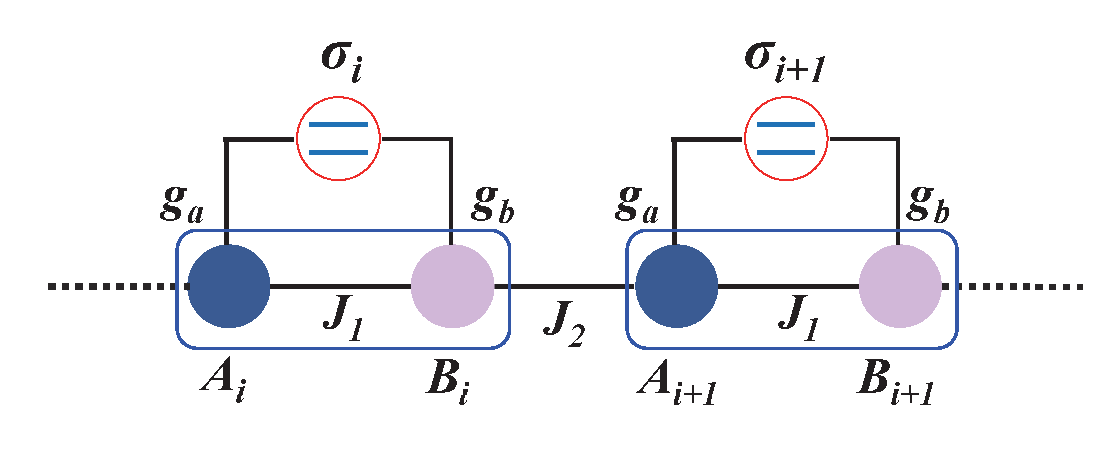}
  \caption{Schematic illustration of the giant atom with transition frequency $\omega_o$ couples to the topological waveguide. The topological waveguide consists of a series of the cell, where each cell contains two sub-lattices $A, B$, and the inter-cell and extra-cell coupling strengths are $J_1, J_2$. The giant atoms with coupling strengths of $g_a$ and  $g_b$ are coupled to sub-lattices $A, B$.  } \label{figure1}
  \end{figure}
 We consider the  giant atoms with multiple coupling nodes coupled to a topological waveguide described by the Su-Schrieffer-Heeger  model. The total Hamiltonian of the system can be written as
\begin{equation}
  \begin{aligned}
    H=H_B+H_o+H_{I}. \label{eq1}
\end{aligned}
\end{equation}
 where
\begin{equation}
\begin{aligned}
  H_B&=\sum\limits_{j=-N}^N\omega_{a} (a_j^{\dagger}a_j+ b_j^{\dagger}b_j)+(J_1a_{j}^{\dagger}b_j + J_2a_{j+1}^{\dagger}b_j+h.c.),\\
 H_o&=\sum\limits_{n=1}^M\omega_{o}\sigma^{\dagger}\sigma, \\
 \quad H_{I}&=\sum\limits_{n=1}^M[(g_{a}a_n^{\dagger}+g_{b}b_{n+l}^{\dagger})\sigma_n+h.c.],
\end{aligned}\label{eq3}
\end{equation}
where $a_{n}^{\dag}$ ($b_{n}^{\dag}$)  and $a_n$ ($b_{n}$)  are the creation and annihilation operators for the cavity.  $\sigma_n=\vert g\rangle\langle  e\vert$ is the spin operators that represents the  transition $\vert e\rangle \to \vert  g\rangle $.   $H_B$ describes the topological waveguide with the SSH  model. As shown in Fig. \ref{figure1}(a), the topological waveguide consists of a series of the cell, where each cell contains two sub-lattices $A$ and  $B$, and the inter-cell and extra-cell coupling strengths are $J_1, J_2$. $H_o$ denotes the free energy of  the giant atom. In addition, in order to unify variables, we set $J_1=J(1-\delta)$,  $J_2=J(1+\delta)$, where we use $J$ as a unit. Moreover, this setting is only used in numerical calculations, whereas in derivations it is still used in $J_1$ an $ J_2$. 
For clarity of description, we will use the topological waveguide frequency  $\omega_a$ as the reference energy, then the free energy of the giant atoms becomes $H_o=\sum\limits_{n=1}^M\Delta \sigma_n^{\dagger}\sigma_n$ with $\Delta=\omega_o-\omega_a$.   $H_{I}$ represents the  interaction between 
 giant atoms  and the topological Waveguide, where the giant atoms with two coupling nodes couples to the sub-lattice $A$ at cell  $n$  and sub-lattice $B$ at cell $n+l$ with coupling strengths $g_{a}$ and $g_{b}$, respectively. 
\begin{figure}[t]
  \centering
  \includegraphics[width=9cm]{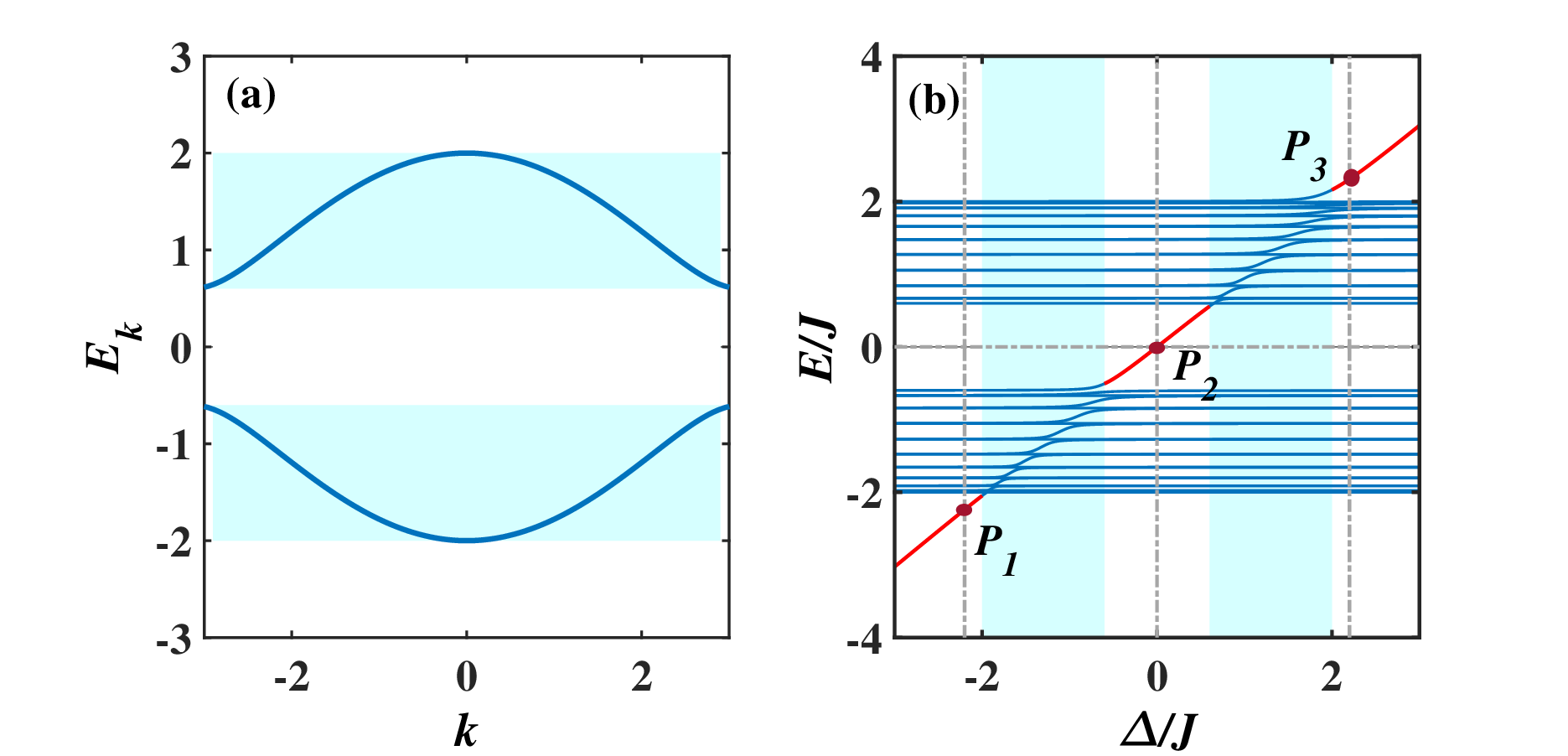}
  \caption{(a) The energy dispersion of the topological waveguide  varies with $k$ in the first Brillouin zone. (b) The energy spectra varies with $\Delta$ for $\delta=0.3$.  The other parameters  $g_a=0.1J,g_b=0.1J$, $m=10$ and $N=20$.} \label{energy}
  \end{figure}
We use the Fourier transform $O_j=\frac{1}{\sqrt{N}}\sum\limits_{k} e^{ikj}O_k$, where $k\in [-\pi,\pi]$. Considering periodic boundary condition, we rewrite $H_B$ and $H_{I}$ as
\begin{equation}
  \begin{aligned}
  H_B&=\sum\limits_k (J_1+J_2e^{-ik})a_k^{\dag}b_k+(J_1+J_2e^{ik})b_k^{\dag}a_k,\\
  &=\sum\limits_k \psi_k^{\dag}H_B(k)\psi_k,
  \end{aligned}
  \end{equation} 
  where $\psi_k^{\dag}=(a_k,b_k)$. The two-band Hamiltonian matrix $H_B(k)$ in momentum space is given by
  \begin{equation}
    H_{B}(k)=\left(
  \begin{array}{ccl}
  0&f(k)\\
  f^{*}(k)&0
  \end{array}
  \right),  \label{eq5}
  \end{equation}%
  where $f(k)=J_1+J_2e^{-ik}=\omega(k)e^{i\phi(k)}$ with $\omega(k)=\sqrt{J_1^2+J_2^2+2J_1J_2\cos(k)}$, $\phi(k)=\arctan\frac{J_2\sin(k)}{J_1+J_2\cos(k)}$.  
  The  dispersion relation of the SSH model can be calculated as $E(k)=\pm \omega(k)$.   
  In Fig. \ref{energy}(a), we plot the energy dispersion  varying with $k$ in the first Brillouin zone. It can be seen that the SSH model can be divided into two parts, also known as energy bands (cyan), in which photons can propagate with velocity $v_g=\partial \omega(k)/\partial k$ and bandgap regions, in which photon propagation is prohibited. 
\begin{equation}
  \begin{aligned}
  H_{I}&=\frac{1}{\sqrt{N}}\sum\limits_{n=1}^M \sum\limits_k[e^{-ikx_n}(g_aa_k^{\dagger}+g_bb_{k}^{\dagger}e^{ilk})\sigma_n+h.c.], \label{eq4}
  \end{aligned}
  \end{equation} 


Given that the SSH model satisfies the chiral symmetry, we can diagonalize $H_B(k)$ as $H_B=\sum\limits_k \omega(k)(\alpha_k^{\dag}\alpha_k-\beta_k^{\dag}\beta_k)$, where the relationship  between the diagonalized basis ($\alpha_k, \beta_k$) and  ($a_k, b_k$) satisfies
\begin{equation}
 \left(
\begin{array}{ccl}
 a_k\\
 b_k
\end{array}%
\right) 
 = 
 \frac{1}{\sqrt{2}}\left(
\begin{array}{ccl}
1&-1\\
e^{-i\phi(k)}&e^{-i\phi(k)}
\end{array}%
\right) 
\left(
  \begin{array}{ccl}
   \alpha_k\\
   \beta_k
  \end{array}%
  \right) .\label{eq6}
\end{equation}%

Substituting (\ref{eq6}) into Eq. (\ref{eq4}), we can obtain 
\begin{equation}
\begin{aligned}
H_{I}=\frac{1}{\sqrt{N}}\sum\limits_{n=1}^M\sum\limits_k\{[p(k)\alpha_k^{\dag}+q(k)\beta_k^{\dag}]e^{-ikx_n}\sigma_n+h.c.\},
\end{aligned}\label{e8}
\end{equation}
where $p(k)=\frac{1}{\sqrt{2}}(g_{a}+g_{b}e^{-ikl}e^{i\phi(k)})$, $q(k)=\frac{1}{\sqrt{2}}(-g_{a}+g_{b}e^{-ikl}e^{i\phi(k)})$. Next we consider two regimes that the bandgap regime and the band regime, which is determined by whether the giant atom frequency falls within the band or not.

 \section{Bandgap regime} 
 \subsection{Topologically Protected giant atom-photon bound state } 
  When the giant atom  frequency lies within the  bandgap, the bound state can form and exponentially localizes in the vicinity of the cavity to which the giant atom is coupled. 
  By solving the Schrödinger's equation $H \ket{\psi_{bs}}=E_{bs}\ket{\psi_{bs}}$ with $E_{bs}$ lying out of the bands,  the energy and wavefunction of bound states in the single excitation subspace can be obtained.   We consider a giant atom couple a cell at $j=m $ and  couple to $A$ and $B$sub-lattice.
  In the single-excitation subspace, the bound state of the Hamiltonian (\ref{e8}) has the form
 \begin{equation}
  \begin{aligned}
  \vert \psi_{bs}\rangle= [C_e\sigma^{\dag}+\sum\limits_{k}(C^\alpha_k\alpha_k^{\dag}+C^\beta_k\beta_k^{\dag})] \vert g,vac\rangle,
  \end{aligned}\label{e9}
  \end{equation}
  where $C_\sigma$ and $ C_k^{\alpha(\beta)}$ are the probability amplitudes for giant atom and photons, respectively. Substituting  the  Hamiltonian (\ref{e8}) and the bound state (\ref{e9}) into  Schrödinger's equation, we can obtain the probability amplitudes of  photons under diagonalized basis   in momentum space 
  \begin{equation}
    \begin{aligned}
     &C^\beta_k=\frac{1}{\sqrt{N}}\frac{C_eq(k)e^{-ikm}}{E_{bs}+\omega(k)},\\
     &C^\alpha_k=\frac{1}{\sqrt{N}}\frac{C_ep(k)q(k)e^{-ikm}}{E_{bs}-\omega(k)}.
    \end{aligned}
  \end{equation}
  And the eigenenergy satisfies the following transcendental equation
     \begin{equation}
      \begin{aligned}
     E_{bs}&=\Delta+\int_{-\pi}^{\pi} \frac{dk}{2\pi}(\frac{\vert p(k)\vert^2}{E_{bs}-\omega(k)}+\frac{\vert q(k)\vert^2}{E_{bs}+\omega(k)}),\\
     &\quad=\Delta+\Sigma(E_{bs}),
    \end{aligned}
  \end{equation}
  where $\Sigma(\cdot)$ is the   self-energy function. 
In fact, it is difficult to obtain an analytical solution of 
the transcendental equation.  We can determine the solution to the equation by theoretically analyzing the behavior of the self-energy function.  Because the $\Sigma(\cdot)$ diverges on all band edges, this ensures that a bound state can be found in each band gap \cite{PhysRevA.93.033833,Bello2019}.
  In addition, we also can  appeal to the numerical diagonalization of the Hamiltonian in real space and plot the energy spectrum of a single excitation under periodical boundary condition  \cite{PhysRevA.101.053855}.   
In Fig. \ref{energy}(b), we plot the energy spectrum varies with $\Delta$.   The energy band marked in blue are the scattering states and the above and below the propagating band marked in red are the bound states. It can be seen that in each bandgap, both have a bound state exist such as points $P_1,P_2,P_3$.  

Combining with the relationship of ($\alpha_k,\beta_k$) and ($a_k,b_k$), we can find that the probability amplitudes for  photons $C^{a(b)}_k$ satisfies
\begin{equation}
  \begin{aligned}
   &C^a_k=\frac{1}{\sqrt{N}}\frac{(g_aE_{bs}+g_b\omega(k)e^{i\phi(k)})e^{-ikm}}{E_{bs}^2-\omega(k)^2},\\
   &C^b_k=\frac{1}{\sqrt{N}}\frac{(g_bE_{bs}+g_a\omega(k)e^{-i\phi(k)})e^{-ikm}}{E_{bs}^2-\omega(k)^2},
  \end{aligned}
\end{equation}
Transferring the probability  amplitudes in momentum space to real space by Fourier transform, we can obtain the probability amplitudes  $C^{a(b)}_j$ in the real space.  Particularly, for $\Delta=0$, we can analysis the probability  amplitudes  in the topological/trivial phase  as shown in Table (\ref{t1}). The detailed derivation in the Appendix \ref{A}. 
\begin{widetext}
\begin{center}
\centering
\begin{table}[h]
\caption{The probability amplitudes  $C^{a(b)}_j$ in the real space for $\Delta=0$.}\label{t1}  
\resizebox{\linewidth}{!}{
\begin{tabular}{|l|l|l|l|l|l|l|}
\hline
~&\multicolumn{3}{c|}{Topological phase}&\multicolumn{3}{c|}{Trivial phases}\\
\hline
~ & \quad $j> m$ & \quad $j=m$ & \quad$j < m$ & \quad$j> m$ & \quad $j=m$ & \quad$j< m$ \\ \hline
$C_j^a$ & \quad $\frac{g_{b}C_e}{J_1}(-\frac{J_1}{J_2})^{j-m}$& \quad\quad 0 &\quad\quad 0 & \quad\quad 0 &\quad $-\frac{g_{b}C_e}{J_1}$ & $-\frac{g_{b}C_e}{J_1}(-\frac{J_1}{J_2})^{(j-m)}$\\ \hline
$C_j^b$ & \quad\quad 0 & \quad\quad 0  &$\frac{g_{a}C_e}{J_1}(-\frac{J_1}{J_2})^{j-m}$ &-$\frac{g_{a}C_e}{J_1}(-\frac{J_2}{J_1})^{j-m}$& \quad $-\frac{g_aC_e}{J_1}$ & \quad\quad 0 \\ \hline
\end{tabular}
}
\end{table}
\end{center}
\end{widetext}

\begin{figure}[h]
\centering
  \includegraphics[width=9cm]{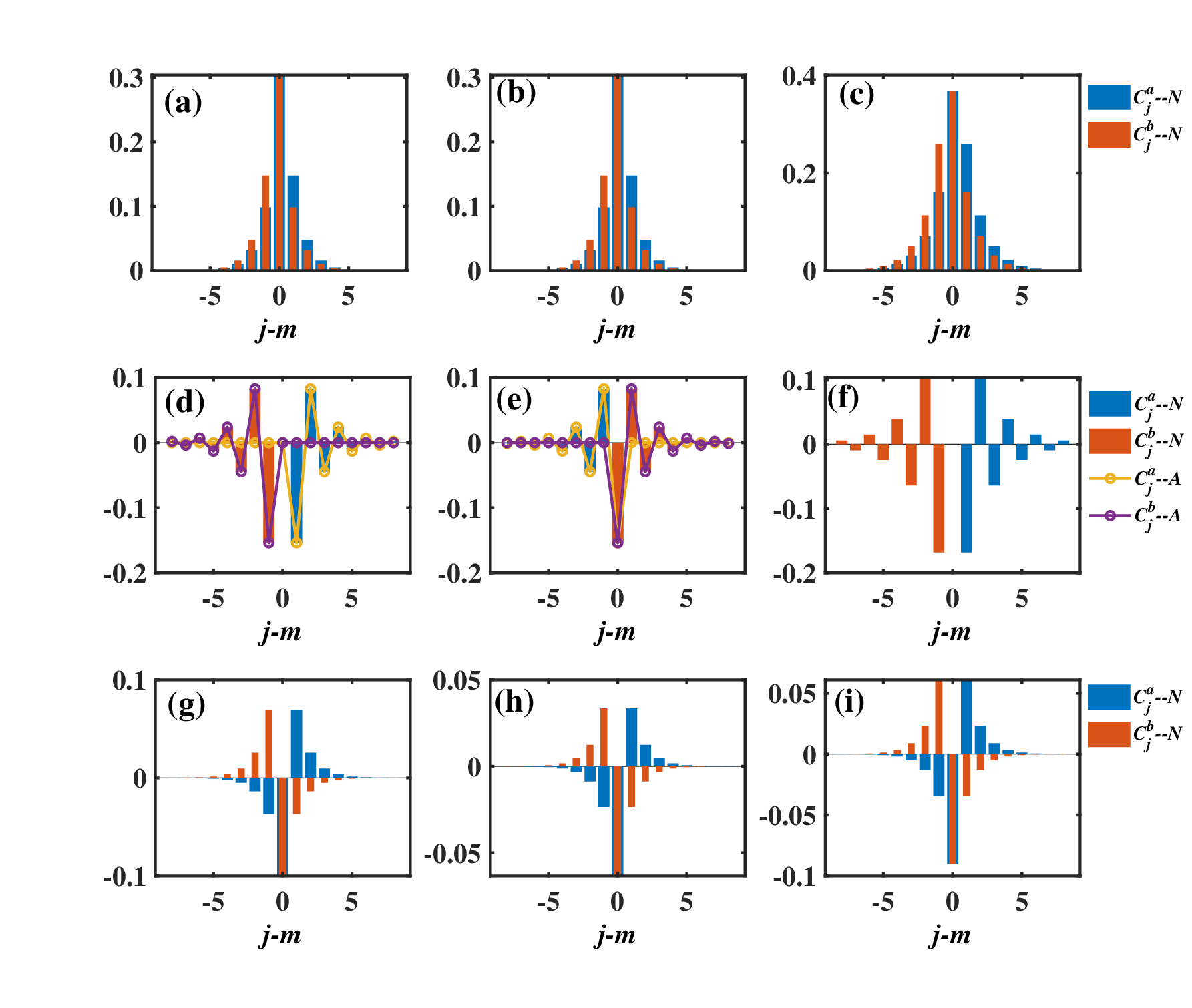}
  \caption{The probability distributions of the bound states for  $\Delta=2.2J$ (a)(b)(c), $\Delta=0J$ (d)(e)(f) and $\Delta=-2.2J$ (g)(h)(i). The first column corresponds to the waveguide in the topological phase, the second column corresponds to the waveguide in the trivial phase and  third column corresponds to the waveguide in the topological phase with off-diagonal disorder and the disorder strength is $0.2$.  
The other parameters  $g_a=0.1J,g_b=0.1J$, $m=10$ and $N=20$.} \label{bs}
  \end{figure}

In Fig. \ref{bs}, we plot the probability  amplitudes  $C_j^{a(b)}$ in real space by numerically calculating Eq. (\ref{e14}).
We select types of detuning corresponding to the three points $P_1,P_2,P_3$  in Fig. \ref{energy}(b).   It can be seen that the photon bound state is exponentially localized in the vicinity of giant atoms.
For $\Delta=\pm2.2J$, we find that the distribution of bound states does not differ in the topological phase from the trivial phase.
For $\Delta=0$,  the wave function amplitudes of photon bound state $C_j^{a(b)}$ in real space   not only depends on the ratio between $J_1$ and $J_2$ but also on the coupling rate of the giant atom-coupled waveguide $g_a,g_b$.   When the strength of the giant atoms coupled to the two sub-lattice points is the same $g_a=g_b$, the probability amplitude of the bound states is symmetrically distributed. If the coupling strength is different $g_a\neq g_b$, it exhibits an asymmetric photon distribution with perfect chirality at $\{g_a,g_b\}=0$, which is consistent with \cite{Bello2019}.
For $\Delta=0$, we find that the photon probability amplitudes have a significant difference for the  topological phase and trivial phase. 
For the topological phase, the photons are distributed only on the sub-lattice $A$ ($B$) with $j\geq m$ ($j\leq m$) and there is no photon distribution at $j = m$. However,  for the trivial phase, the result is the opposite as shown in Table (\ref{t1}).
As shown in Figs. \ref{bs}(d)(e),  we can see that  the numerical results are in agreement with the theoretical results.
For $\Delta=0$, the giant atom-photon bound state can be viewed as a topological edge mode, which can inherit the information from the topological waveguide, e.g. robustness to the off-diagonal disorder\cite{Bello2019}. We can find the  slight off-diagonal  disorder does not affect the possibility  amplitudes of the photon bound state. However, for $\Delta=\pm 2.22J$,  the distribution of the probability amplitude is sensitive to off-diagonal disorder as shown in Figs. \ref{bs}(c)(f)(i).

\subsection{Virtual photon induced dipole interactions between giant atoms} 
In the bandgap, photons are forbidden to propagate. In this case, the dipole interactions between the giant atoms originate from the overlap of the photon bound states. By adiabatically eliminating photon modes, we can obtain dipole interactions between giant atoms as 
\begin{equation}
\begin{aligned}
  H_{S}=& \sum\limits_{n,m}(J_{nm}\sigma_n^\dag \sigma_m+h.c).
\end{aligned} \label{hs}
\end{equation}
with $J_{nm}=g_ag_b\frac{(-1)^{x_{nm}}}{J_1}(\frac{J_1}{J_2})^{x_{nm}}$ for $x_{nm}\geq 1$ and the waveguide is in the topological phase. See Appendix \ref{C} for details.
\begin{figure}[t]
\centering
\includegraphics[width=9cm]{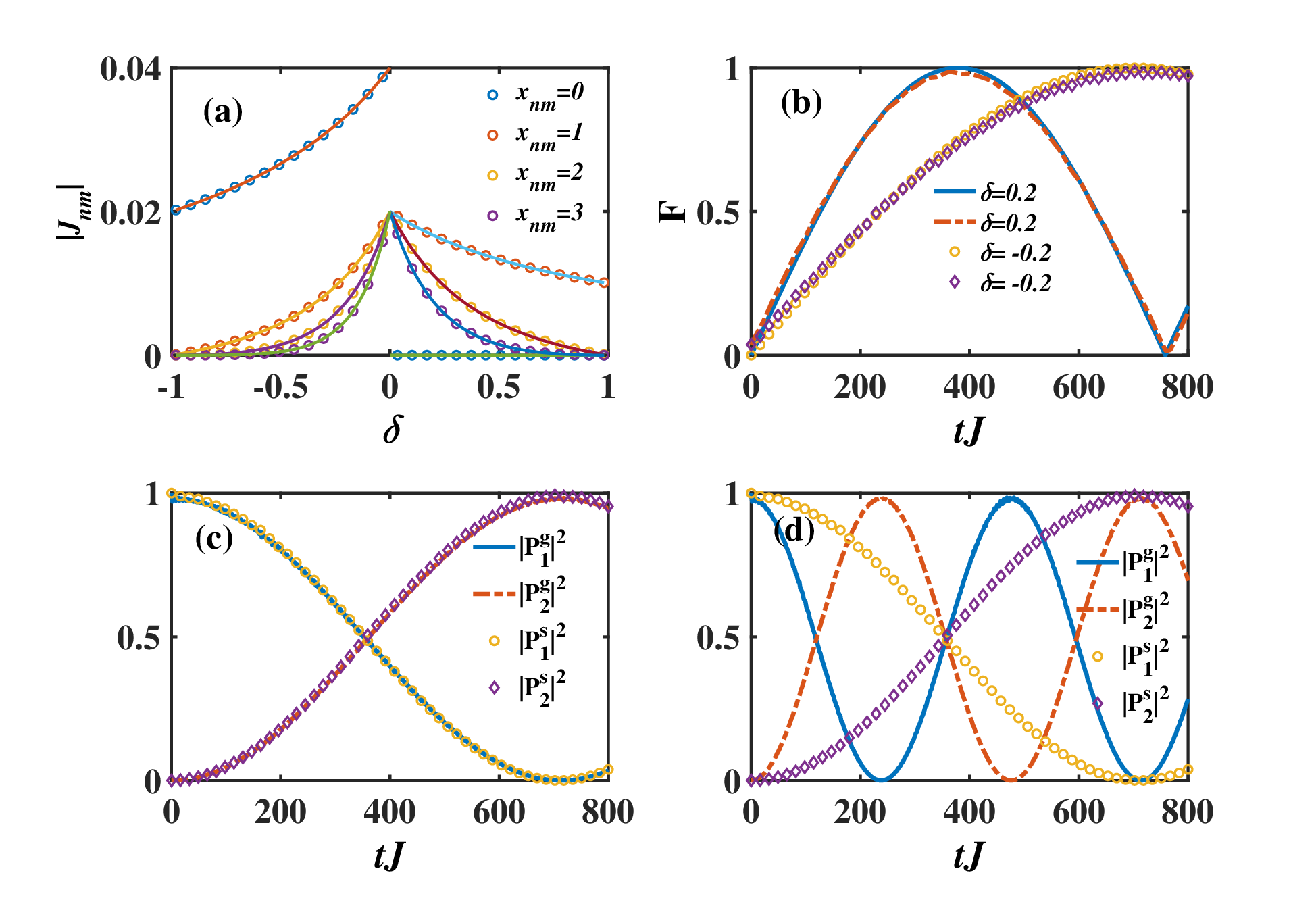}
\caption{(a) The coherent interaction $J_{nm}$ between giant atoms  as a function of position $x_{nm}$. (b) The  fidelity $F$ varying with  $t$ for $\delta=\pm0.2$ governed by Eq. (\ref{eq1}) the dashed line and  Eq. (\ref{hs}) the  blue line.  (c)(d) The excitation transfer of small ($|p^s_1|^2,|p^s_2|^2$) and giant atoms ($|p^g_1|^2,|p^g_2|^2$) changes in time with $\delta=0.5J$. The other parameters  $g_a=0.1J,g_b=0.1$, $N=40$ and $\Delta=0$.
} \label{figure222}
\end{figure} 
In Figs. \ref{figure222}(a), we plot the long-range coherent interaction between giant atoms  $J_{nm}$ varying with $x_{nm}$. We can find that when $x_{nm}=0$ namely only one giant atom is coupled to the waveguide, the frequency  of giant atoms exists energy shifted in the  trivial  phase, while in the  topological phase, the giant-atom frequency is not affected.
For $x_{nm}\neq0$, the coherent  interactions $J_{nm}$ gradually increase as $\delta$ increases, while in the topological phase, it decreases as $\delta$ increases except for $x_{nm}=0$. 
When $\delta=0.9J$ $(-0.7J)$,  only near-neighbor interactions will exist with other long-range interactions approximated to zero in the topological (trivial) phase. For $|\delta|<0.5$, it allows for the existence of more long-range interactions, which can be used to investigate the spin models with long-range interactions \cite{PhysRevLett.130.213605,Richerme2014} and quantum simulations \cite{Douglas2015,PRXQuantum.3.010336}.
Moreover, we also find that the coherent interactions induced in the topological phase are  larger than
that in the  trivial phases under the same
band gap width such as $\delta=\pm 0.2$. 

Next, we consider two giant atoms coupled to different cells, where two coupling nodes of each giant atom are coupled in two sub-lattices of a cell. And the first giant atom is initially in the excited state and the other giant atoms are in the ground state. 
By solving the Schrödinger's equation governed by Hamiltonian (\ref{hs}), we can theoretically calculate the  population evolution of the two giant atoms is $P_1=cos(J_{nm}t)$ and $P^2=sin(J_{nm}t)$ respectively.    
In  Fig. \ref{figure222}(b), we plot the  fidelity $F$ varying  $t$ by  numerically solving Eq. (\ref{eq1}) and  Eq. (\ref{hs})  for $\delta=\pm0.2$.  The fidelity is defined by $F=\langle \psi_t| \psi_f\rangle$ with $|\psi_f\rangle$ is the  wave function at time $t$ and $|\psi_t\rangle$ represents  the target state, where the  second giant atom is in the excited state as well as the first in the ground state.
We consider the distance between two giant atoms are $x_{nm}=2$, and the waveguide is in the topological or trivial phase. 
It can be found that the period of fidelity is smaller in the topological phase than in the trivial phase.  This means that the excitation transfer of giant atoms is faster in the topological phase than in the trivial phase.  

In addition, we also consider that giant atoms are replaced by small atoms.  The spacing and coupling between two small atoms is the same as in the case of giant atoms, but excitation transfer can only be realized if the first atom is coupled to sub-lattice $A$ and the second atom to sub-lattice $B$ when the waveguide in the Topological phase \cite{Bello2019,PhysRevX.6.021027}.
In Fig. \ref{figure222}(c), we plot the populations of  giant atoms ($|p^g_1|^2,|p^g_2|^2$) and small atoms ($|p^s_1|^2,|p^s_2|^2$) varying with time. 
It can be clearly seen that the behavior of the evolution of small atoms is equivalent to that of giant atoms. 
When we change the coupling positions of the giant atoms, e.g., the first coupling point of the giant atoms to the $j$th sub-lattice $A$, and the second coupling point to the $(j+1)$th sub-lattice $B$, we find that the excitation transfer period of the giant atoms at this time is twice that of the small atoms as shown in Fig. \ref{figure222}(d). This means that giant atoms can achieve faster quantum transmission. 
\section{Band regime}
\subsection{Photon-induced coherent and dissipative coupling between giant atoms.} 

    In this section,  we derive the master equation which governs the dynamics of the system by considering the topological waveguide  as the structured environment. In the interaction picture, the interaction Hamiltonian (\ref{e8}) is written as 
    \begin{equation}
      \begin{aligned}
      H_{int}(t)&=\sum\limits_{n=1}^M\sum\limits_k\{e^{-ikx_n}[p(k)e^{i\omega(k)t}\alpha_k^{\dag}\\
      &\quad+q(k)e^{-i\omega(k)t}\beta_k^{\dag}]\sigma_ne^{-i\Delta t}+h.c.\}.
      \end{aligned}
    \end{equation}
    If coupling between giant atoms and topological waveguides can be treated with the Born-Markov approximation, then it is possible to obtain a master equation that effectively describes the dynamics of giant atoms by tracking photon degrees of freedom \cite{10.1093/acprof:oso/9780199213900.001.0001}. When the giant atoms are resonant with one of the bands $\Delta \in [ -\omega(k), \omega(k)]$, then the reduce master equation is given by 
    \begin{equation}
      \begin{aligned}
      \frac{d\rho }{dt}=\sum\limits_{n,m=1}^M\Gamma_{nm}(\sigma_n\rho\sigma^{\dag}_m-\sigma^{\dag}_m\sigma_n\rho)+h.c.,
      \end{aligned}
    \end{equation}
    with the reservoir-mediated coupling between giant atoms

    \begin{equation}
      \begin{aligned}
     \Gamma_{nm}&=\lim\limits_{\eta\to0^+}\sum\limits_{\substack{k\\\beta=l,u}}\frac{\langle 0\vert\sigma_n H_{int}\beta_k^\dag\vert 0\rangle\langle 0\vert \beta_kH_{int}\sigma_m^\dag \vert 0\rangle}{\eta-i(\Delta-\omega_\beta(k))}e^{ikx_{nm}},\\
       &=\lim\limits_{\eta\to0^+}\sum\limits_{k}\frac{1}{N}[\frac{\vert p(k)\vert^2e^{ikx_{nm}}}{\eta-i(\Delta-\omega(k))}
       +\frac{\vert q(k)\vert^2e^{ikx_{nm}}}{\eta-i(\Delta+\omega(k))}],\\
    \end{aligned} \label{eq24}
    \end{equation}
    where $\omega_u(k)=-\omega_l(k)=\omega(k)$, $x_{nm}=x_n-x_m$ and $\vert 0 \rangle$ is the ground state of the system.
    We can separate the real and imaginary contributions of $\Gamma_{nm}$,  then we can obtain 
    \begin{equation}
      \begin{aligned}
      \frac{d\rho }{dt}&=-i\sum\limits_{n,m}J_{nm}[\sigma^{\dag}_m\sigma_n, \rho]\\
      &\quad + \sum\limits_{n,m}\gamma_{nm}(2\sigma_n\rho\sigma^{\dag}_m-\sigma^{\dag}_m\sigma_n
        \rho-\rho\sigma^{\dag}_m\sigma_n),
      \end{aligned}
    \end{equation}
    where $J_{nm}=Im(\Gamma_{nm})$ and  $\gamma_{nm}=Re(\Gamma_{nm})$ are the  coherent and dissipation interactions mediated by the topological waveguide, respectively.
 Next, we set $z=\Delta+i\eta$, the  Eq. (\ref{eq24}) becomes
\begin{equation}
\begin{aligned}
     \Gamma_{nm}(z)&=i\int_{-\pi}^{\pi} \frac{dk}{2\pi}(\frac{\vert p(k)\vert^2}{z-\omega(k)}+\frac{\vert q(k)\vert^2}{z+\omega(k)})e^{ikx_{nm}}.\\
\end{aligned} \label{ga}
\end{equation}   

\begin{figure}[t]
\centering
\includegraphics[width=9cm]{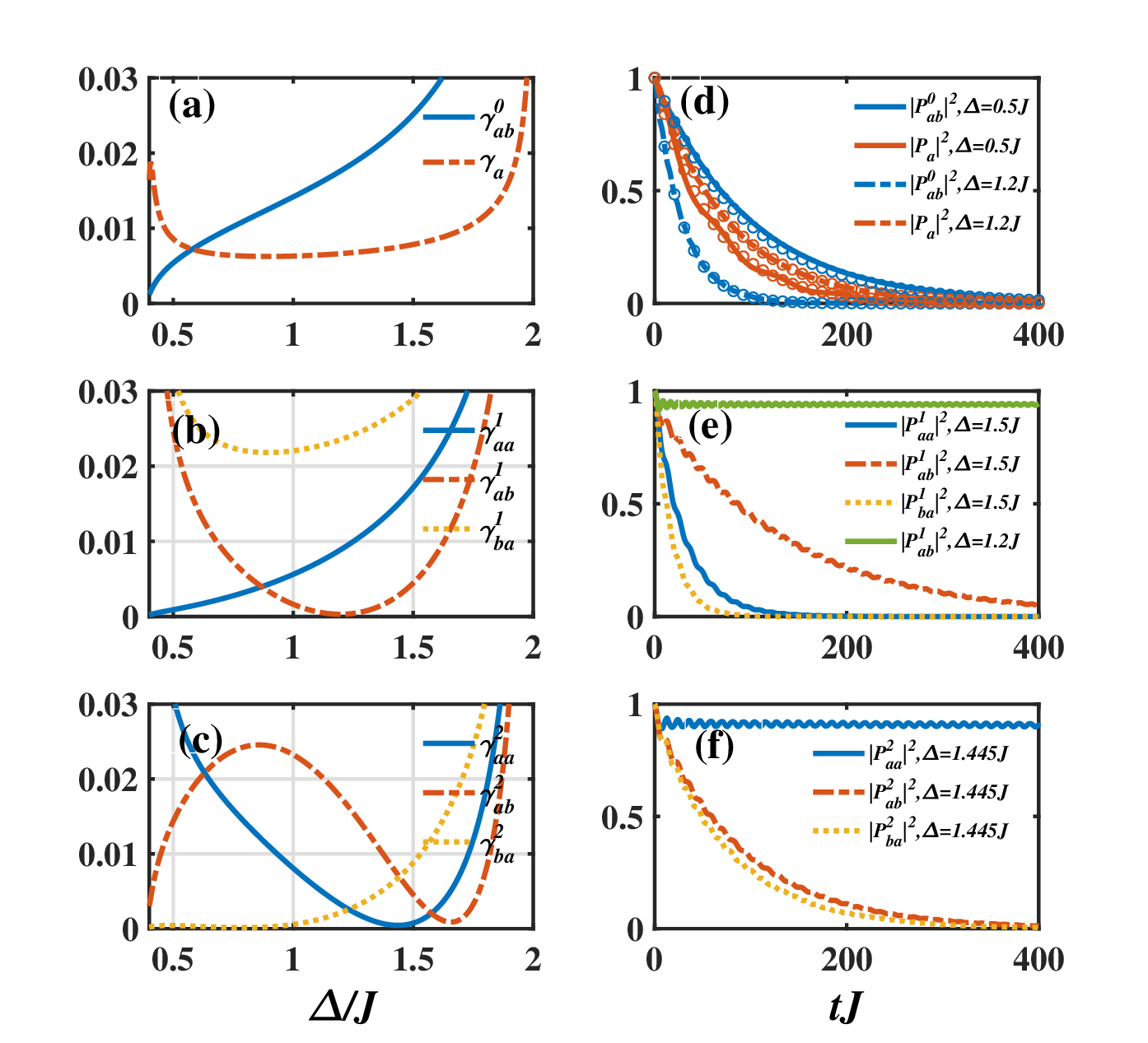}
\caption{ (a)-(c)The decay rate of giant atoms under Markov approximation varies with  $\delta$ for different $l$. (d)-(f) The probability of giant atom governed by Hamiltonian (\ref{eq1}) varies with time for different coupling nodes corresponding to the left figures.  The figure form top to bottom  for the distances between the coupling nodes $l=0,1,2,3$.  The circles represent the evolution of  small  atom and giant atom under the Markov approximation.
The parameters are $ \{g_{a0}, g_{b0}, g_{al}, g_{bl}\}=0.1J$.
    } \label{GE}
\end{figure}
Next, we consider a single giant atom coupling to the topological with two coupling nodes.

In this case, there are several coupling scenarios: (I) Two nodes in one cell coupled to sub-lattices $A$ and $B$. (II) Two nodes in different cells, one node coupled to sub-lattice $A$ or $B$ in cell $j=m$ and the other node coupled to sub-lattice $A$ or $B$ in cell $j=m+l$.
For the above two cases, by appropriately selecting the parameters, the coupling can be written uniformly in the following form as $p(k)=\frac{1}{\sqrt{2}}[(g_{a0}+g_{al}e^{-ikl})+(g_{b0}+g_{bl}e^{-ikl})e^{i\phi(k)}])$, $q(k)=\frac{1}{\sqrt{2}}[-(g_{a0}+g_{al}e^{-ikl})+(g_{b0}+g_{bl}e^{-ikl})e^{i\phi(k)})]$. We note that when $\{g_{al},g_{bl},g_{b0}\}=0$ the model reduces to the small atom coupling to the topological waveguide, when $\{g_{al},g_{bl}\}=0$ or $\{g_{a0},g_{b0}\}=0$ the model reduce to the case (I), and when $\{g_{a0},g_{bl}\}=0$ or $\{g_{al},g_{b0}\}=0$ or  $\{g_{a0},g_{al}\}=0$ or $\{g_{b0},g_{bl}\}=0$ the model reduce to the case (II). In fact, $\{g_{a0},g_{b0}\}=0$ and $\{g_{al},g_{bl}\}=0$ ($\{g_{a0},g_{al}\}=0$ and $\{g_{b0},g_{bl}\}=0$)  are equivalent in geometric configuration.  Therefore, four forms of coupling exist for a single giant atom coupled to the topological waveguide with two nodes. 


Next, we focus on the decay rate $\gamma=Re[\Gamma_{nm}]$.  In Fig. (\ref{GE}), we plot the decay rate of giant atom for the four forms of coupling varying with $\Delta$ by numerically solving the Eq. (\ref{eq24}) and the corresponding evolution of giant atoms with time governed by Hamiltonian (\ref{eq1}). The diagram corresponds from top to bottom to the distances between the coupling nodes $l=0,1,2$. We use $\gamma_{oo^\prime}^l$  to mark the  giant atomic decay rate, where $o,o^\prime=a, b$ represent the coupled sub-lattice and $|P_{oo^\prime}^l|^2$ to represent the evolution of giant atom over time.

For the small-atom case, the decay rate of the atoms $\gamma_a$ is a U-shaped distribution that varies with detuning. However, for the giant atom, the  U-shaped decay distribution vanishes and the decay $\gamma^0_{ab}$ increases gradually as $\Delta$ increases and lager than $\gamma_a$ except for $\Delta \in [0.4,0.56]$. When $\Delta=1.5J$,  the small atom (giant atom) decay is approximately equal to $\gamma_a\approx 0.004J$ ($\gamma_a\approx 0.0125J$) as shown in Fig. \ref{GE}(a). The corresponding to  the evolution of the   small atom (giant atom) under this case is $|P_a|^2=e^{-2\gamma_a t}$ ($|P^0_{ab}|^2=e^{-2\gamma^0_{ab} t}$).
\begin{figure}[t]
  \centering
  \includegraphics[width=9cm]{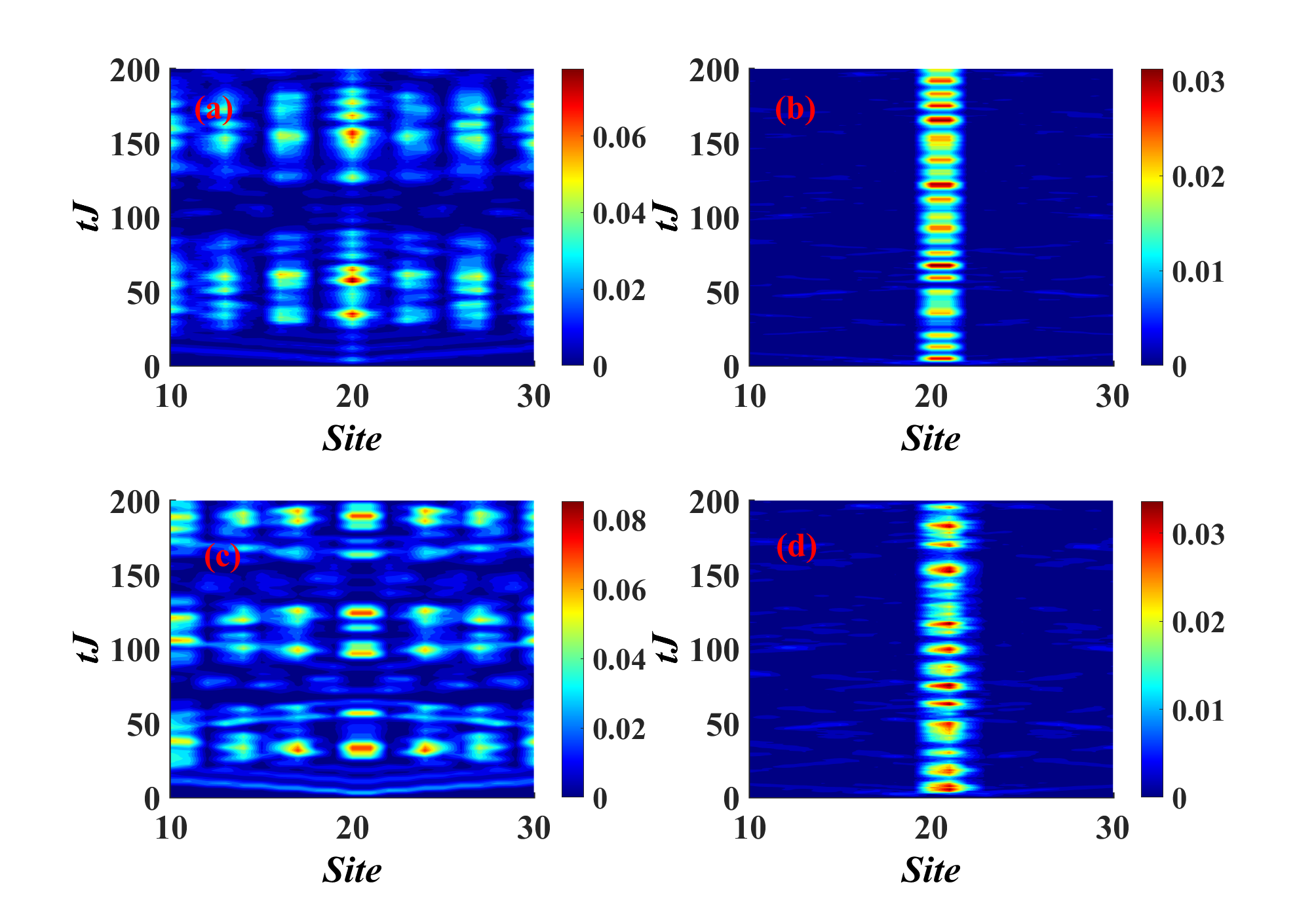}
  \caption{The evolution of the average photon number $n_{o,j}$.  (a)(b)(c) The coupled form of the giant atom $\{g_{a0},g_{al}\}$, $\{g_{a0},g_{bl}\}$ and $\{g_{b0},g_{al}\}$, $l=1$ and $\Delta=1.2J$.   (d)The coupled form of the giant atom $\{g_{a0},g_{al}\}$, $l=2$ and $\Delta=1.445J$. The parameters are $ \{g_{a0}, g_{b0}, g_{al}, g_{bl}\}=0.1J$.   } \label{PE}
  \end{figure}
In  Fig. \ref{GE}(d), we plot the evolution of the giant atom and small atom governed by Hamiltonian (\ref{eq1}) marked in line and under Markov approximation marked in circle. 
We can see that when $\Delta=0.5J$, the excitation of small atom decay faster than that of giant atom and when $\Delta=1.2J$, the excitation of small atom decay faster than that of giant atom.
Furthermore, we find that when $\Delta$ tends to the edge of the energy band, the evolution under the Markov approximation is different from that governed by the Hamiltonian (\ref{eq1}).
We can see that there are fluctuations in the actual evolution process, which means that the system evolution corresponds to a non-Markovian dynamical process.  When $\Delta$ lie in  the center of the energy band, the dynamics predicted by the Markov approximation are consistent with the actual evolution, so we can use the results of the Markov approximation to  characterize the current system.

In Figs. \ref{GE}(b)(c), we plot the the case (II) with $l=1,2 $ for the three different coupling situations. The corresponding evolution of the giant atom  governed by Hamiltonian (\ref{eq1}) are shown in Figs. \ref{GE}(e)(f).  We find  when $\Delta=1.2J$, $\gamma^1_{ab}=0$ for $l=1$, and when $\Delta=1.445J$, $\gamma^2_{aa}=0$ for $l=2$. This implies that in this configuration the evolution of giant atom will be a non-decaying behavior.  The evolution of the giant atom  governed by Hamiltonian (\ref{eq1}) under this configuration are shown in Figs. \ref{GE}(e)(f), marked in green line and blue line, respectively.  This non-decaying oscillatory behavior often corresponds to the appearance of bound states in non-Markovian dynamical processes.
In Figs. \ref{PE} (a)-(c), we plot the dynamics of the average photon number $n_{o,j}=\langle C_{o,j}^{\dag}C_{o,j}\rangle$ with $o=A,B$ on the site for $l=1$.
For the coupled form of the giant atom $\{g_{a0},g_{al}\}$ and $\{g_{b0},g_{al}\}$, as show in  Figs. \ref{PE} (a) and(c),
the photons initially are distributed only in the giant atomic coupling positions and rapidly disperse throughout the waveguide as time evolves.
However, for $\{g_{a0},g_{al}\}$, the photon are trapped between two coupling points of the giant atom,  and as time evolves the trapped photons continue to leak into the waveguide. 
For $l=2$, we  also find that the photons are trapped between the two coupling points and do not diffuse into the waveguide as shown in Fig. \ref{PE} (d). 
Furthermore, for a single giant atom with multiple coupling nodes coupled to the waveguide, the interference effects will be richer and meet the non-dissipative detuning points will be increased. We will not discuss this here.  

\subsection{Quantum state transfer based on giant atom interference effects}
In this section, we study the quantum state transfer based on interference effects for two giant atoms. We find that for the two giant atoms with distance $l^\prime$, the correlated dissipation mediated by the waveguide and the  decay can be suppressed by selecting the appropriate form of coupling, simultaneously,  the coherent interaction remains so that we can realize quantum state transfer.

\begin{widetext}

  \begin{figure}[h]
    \center
    \includegraphics[width=18cm]{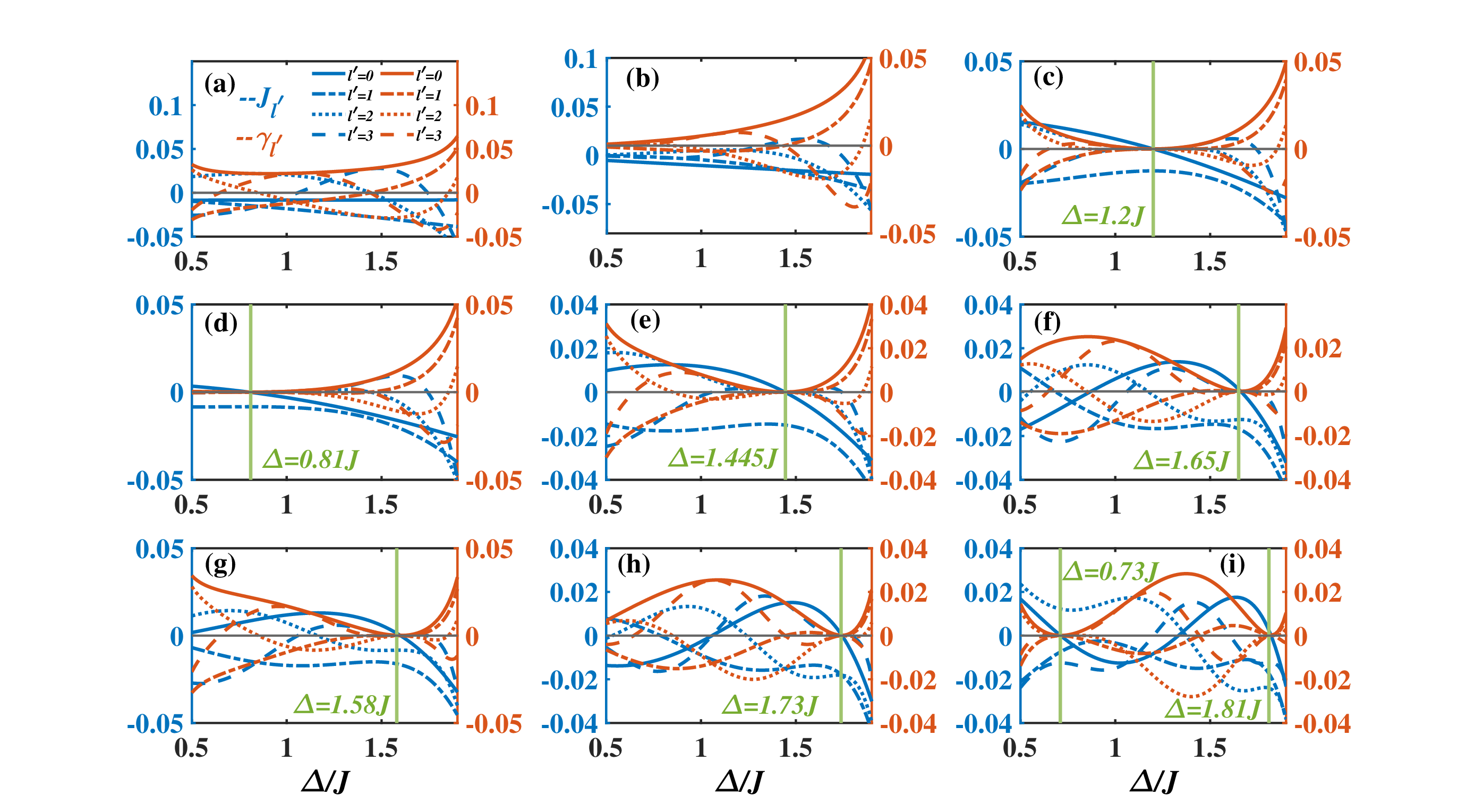}
    \caption{The coherent interaction $J_{l^\prime}$  and correlated dissipation $\gamma_{l^\prime}$ varying with $\Delta$ and $l^\prime=0,1,2,3$. (a)(b)(c) The coupled form of the giant atom $\{g_{b0},g_{al}\}$, $\{g_{a0},g_{al}\}$ and $\{g_{a0},g_{bl}\}$. The graphs correspond from top to bottom to $l=1,2,3$. The blue line represents coherent coupling and the orange line represents correlated dissipation. The green line represents waveguide-induced spontaneous radiation, and correlated dissipation to zero.  The parameters are $ \{g_{a0}, g_{b0}, g_{al}, g_{bl}\}=0.1J$.
    } \label{Two}
  \end{figure} 
  
  \end{widetext}

We consider each giant atom with two coupling nodes as case (II): $\{g_{b0},g_{al}\}$, $\{g_{a0},g_{al}\}$ and $\{g_{a0},g_{bl}\}$.
In Fig. \ref{Two}, we plot the coherent interaction $J_{l^\prime}$ and correlated dissipation $\gamma_{l^\prime}$ varying with $\Delta$ for   case (II).  For $l=1$,   we find that the decay and correlation dissipation are both zero only if the giant atom coupling form satisfies$\{g_{a0},g_{bl}\}$ and $\Delta=1.2J$, but the coherent coupling is non-zero just when $l^\prime=3$ as shown in Fig. \ref{Two}(c).  
When $l>1$, it can be seen that all present appropriate detuning marked in green line to satisfy that the decay and correlation dissipation are zero and the coherent coupling is not zero. The coherent coupling does not necessarily decrease with increasing distance, but may also be the same as shown in Fig. \ref{Two}(h) when  $\Delta=1.73J$, which is different from the bandgap regime.
In particular, for coupled form of the giant atom $\{g_{b0},g_{al}\}$ and $\{g_{a0},g_{al}\}$,  when $l=2$, we find that coherent coupling is not zero only when $l^\prime=1$; when $l=3$, we find that coherent coupling is not zero when $l^\prime=1,2$. We note that the spacing between the giant atoms in this case is less than the distance between the two coupling points of the giant atoms, i.e., the bridge structure. Thus for two giant atoms coupled in a bridge-type structure into a topological waveguide, it is possible to realize the decay and correlation dissipation are zero and the coherent coupling is not zero. 

\begin{figure}[t]
\centering
\includegraphics[width=9cm]{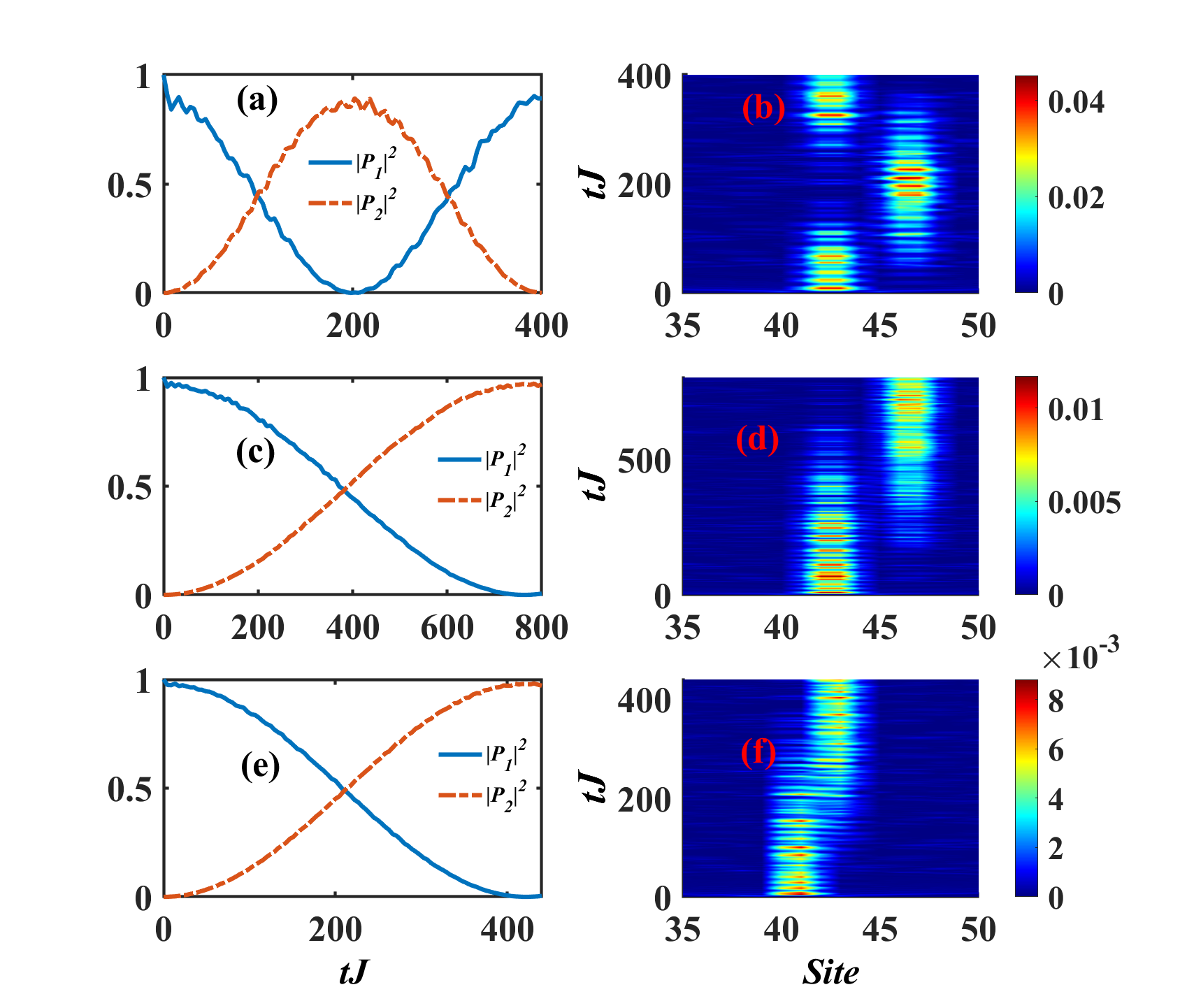}
\caption{Excitation transfer of giant atoms and the  evolution of the average photon number. (a)(b) The coupled form of the giant atom $\{g_{b0},g_{al}\}$ for $l^\prime=2, l=3$ with  $\Delta=1.745J$. (e)The coupled form of the giant atom $\{g_{a0},g_{al}\}$ for $l^\prime=1, l=2$  with  $\Delta=1.445J$. (c)(d)(f) the the average photon number vary with time on the site corresponding to (a)(c)(e), respectively.
The parameters are $\{g_{b0},g_{al}\}=0.1J$ for (a)(b), $\{g_{b0},g_{al}\}=0.05J$ for (c)(d), $\{g_{a0},g_{al}\}=0.05J$ for (e)(f) and $N=40$.} \label{Two_evo}
\end{figure}

The above results means that we can achieve the transmission of quantum information. Next, we consider the first giant atom is in the excited state and the second  giant atom is in the ground state. 
In Fig. \ref{Two_evo} (a), we consider the  coupled form of the giant atom $\{g_{b0},g_{al}\}=0.1J$ for $l^\prime=2, l=3$ with  $\Delta=1.745J$.
It can be seen that the excitation is transferred from the first giant atom to the second giant atom with the  period roughly $T=200/J$, but with only a $90\%$ success rate. This is due to information leakage into the waveguide as shown in Fig. \ref{Two_evo} (b). We can see that the photon is localized between the two giant-atom coupling points and oscillates with time. Because the photons do not diffuse throughout the waveguide therefore there is no loss with the whole state transfer process, in which  corresponding would be a two-giant-atom photon bound state.
When we reduce the coupling strength of the giant atoms to the waveguide, we find that the probability of successful excitation transfer is up to $97\%$, the period is doubled as shown in Fig. \ref{Two_evo} (c).
For other case, e.g. the  coupled form of the giant atom  $\{g_{a0},g_{al}\}$ for $l^\prime=1, l=2$  with  $\Delta=1.445J$,  The giant atom evolution over time and  the distribution of photons are shown in Figs. \ref{Two_evo} (e)(f).

\section{ conclusion}

In this paper, we consider giant atoms coupled to a one-dimensional topological waveguide reservoir. We studied the following two cases.
In the bandgap regime, where the giant-atom frequency lies outside the band, we study the generation and distribution of giant atom-photon bound states and the difference between the topological waveguide in topological and trivial phases.   When the strengths of the giant atoms coupled to the two sub-lattice points are equal, the photons distribution is symmetrical and the  chiral photon distribution is exhibited when the coupling is different. The coherent interactions between giant atoms are induced by virtual photons, or can be understood as an overlap of photon bound-state wave functions, and decay exponentially with increasing distance between the giant atoms. We also find that the coherent interactions induced by the topological phase are larger than those induced by the trivial phase for the same bandgap width.
In the band regime, the giant-atom frequency lies in the band, under the Born-Markov approximation, we obtained effective coherence and correlated dissipative interactions  between  the giant atoms mediated by topological waveguide reservoirs, which depend on the giant-atom coupling nodes.  
We analyze the effect of the form of the giant-atom coupling point on the decay, and on the associated dissipation.  The results show that we can design the coupling form as well as the frequency of the giant atoms to achieve zero decay and correlation dissipation and non-zero coherent interactions.  Finally we used this scheme to realize the excitation transfer of giant atoms.
.

\section{acknowledgments}
This work is supported by NSFC under Grants No.12274053 and National Key R$\&$D Program of China (No.2021YFE0193500).

\begin{appendix}
\section{ Probability amplitudes in the real space}\label{A}
Transferring the probability  amplitudes in momentum space to real space by Fourier transform, we can obtain the probability amplitudes  $C^{a(b)}_j$ in the real space 
\begin{equation}
  \begin{aligned}
   &C^a_k=\frac{1}{\sqrt{N}}\frac{(g_aE_{bs}+g_b\omega(k)e^{i\phi(k)})e^{-ikm}}{E_{bs}^2-\omega(k)^2},\\
   & C^b_k=\frac{1}{\sqrt{N}}\frac{(g_bE_{bs}+g_a\omega(k)e^{-i\phi(k)})e^{-ikm}}{E_{bs}^2-\omega(k)^2},
  \end{aligned}\label{A1}
\end{equation}
If  the  giant atomic frequency resonates with the frequency of the waveguide $\Delta=0$, we can chooses the energy of bound state $E_{bs}=0$  shown in Fig. \ref{bs} (b), the above integral  can be reduced 
\begin{equation}
  \begin{aligned}
   &C^a_j
   =-g_b\int_{-\pi}^{\pi} \frac{dk}{2\pi}\frac{(J_1+J_2e^{-ik})e^{ik(j-m)}}{J_1^2+J_2^2+2J_1J_2\cos(k)}, \\
   &
   C^b_j
   =-g_a\int_{-\pi}^{\pi} \frac{dk}{2\pi}\frac{(J_1+J_2e^{ik})e^{ik(j-m)}}{J_1^2+J_2^2+2J_1J_2\cos(k)}.
  \end{aligned}\label{e14}
\end{equation}
Define $y=e^{ik}$,  the domain of integration is now the unit circle on the complex plane. Then, the above  integral Eq. (\ref{e14}) becomes
\begin{subequations}
  \begin{align}
   &C^a_j=-\frac{dk}{2\pi i}\oint\limits_{\vert y\vert=1}\frac{g_{b}y^{j-m-1}dy}{(J_2y + J_1 ) }, \\
   &C^b_j=-\frac{dk}{2\pi i}\oint\limits_{\vert y\vert=1}\frac{g_{a}y^{j-m}dy}{(J_2 + J_1y) }.
  \end{align}
\end{subequations}
For $j\geq m$, applying  Cauchy's residue theorem, we can obtain 
\begin{equation}
C_j^a=\left\{
\begin{array}{rcl}
      &\frac{g_{b}}{J_1}(-\frac{J_1}{J_2})^{j-m},& J1<J2;\\
      &0, &J1>J2;
\end{array}%
      \right.
    \end{equation}%
    \begin{equation}
C_j^b=\left\{
\begin{array}{rcl}
      &0,& J1<J2;\\
      &-\frac{g_{b}}{J_1}(-\frac{J_2}{J_1})^{j-m}, &J1>J2;
\end{array}%
      \right.
\end{equation}%

The same steps for $j\leqslant m$ and trivial phase, we can obtain Table \ref{t1}.

  \section{Dipole-dipole interaction by adiabatically eliminating photon modes}
  \label{C}
Under the diagonal basis vector, the complete Hamiltonian is as follows  
\begin{equation}
\begin{aligned}
H=H_0+H_I,
\end{aligned}
\end{equation}
with the free Hamiltonian $H_0$ and the interaction Hamiltonian $H_I$.
\begin{equation}
\begin{aligned}
H_{0}&=\sum\limits_k \omega(k)(\alpha_k^{\dag}\alpha_k-\beta_k^{\dag}\beta_k)+\sum\limits_{n=1}^M\omega_{o}\sigma_n^{\dagger}\sigma_n,\\
H_{I}&=\frac{1}{\sqrt{N}}\sum\limits_{n=1}^M\sum\limits_k\{e^{-ikx_n}[p(k)\alpha_k^{\dag}+q(k)\beta_k^{\dag}]\sigma_n+h.c.\},
\end{aligned}
\end{equation}
Next, we use Schrieffer-Wolff transformation to adiabatically eliminated the photon modes \cite{PhysRev.149.491}.  Define $S=\sum\limits_{k}\lbrace-(\eta_{u,k} \alpha_k+\zeta_{l,k} l_{k})\sigma^{\dag}_n+(\eta^*_{u,k} u^{\dag}_k+\zeta^*_{u,k} l^{\dag}_{k})\sigma_n\rbrace $, where $S$ is an anti-Hermitian operator  $S^{\dag}=-S$ \cite{PhysRev.149.491}. Then,  after using Schrieffer-Wolff transformation, the  Hamiltonian becomes 
\begin{equation}
  \begin{aligned}
  H_{S}=UHU^{\dag}=H+[S, H]+\frac{1}{2}[S,[S, H]]+\cdots,
  \end{aligned}
  \end{equation}
where $U=\exp(S)$. By setting the term  of the first-order perturbation to zero as $H_I+ [H_0,S]=0$
, we can find that the  determined parameters satisfies
$\eta_{u,k}=\frac{p(k)e^{-ikx_n}}{\sqrt{N}(\Delta-\omega(k))}$, $\zeta_{l,k}=\frac{q(k)e^{-ikx_n}}{\sqrt{N}(\Delta+\omega(k))}$. Only \{ $\eta_{u,k}$, $\zeta_{l,k}$ \} $\ll 1$, we can keep only the second-order terms and safely omit the higher order terms. This is equivalent to the large detuning condition as $\{g_a,g_b\}\ll |\Delta-\Delta_{edge}|$, where $\Delta_{edge}$ is the band edge as $\{-2J, -2J\delta, 2J\delta, 2J\}$.
 Then,  under the second-order perturbation approximation, we can obtain 
\begin{equation}
\begin{aligned}
  H_{S}=& \sum\limits_{n,m}(J_{nm}\sigma_n^\dag \sigma_m+h.c).
\end{aligned}
\end{equation}
with $J_{nm}=\sum\limits_{k}(\frac{\vert p(k)\vert^2}{\sqrt{N}(\Delta-\omega(k))}+\frac{\vert q(k)\vert^2}{\sqrt{N}(\Delta+\omega(k))})e^{ikx_{nm}}$. By replacing the discrete modes by the continuous distribution, we can obtain
\begin{equation}
      \begin{aligned}
     J_{nm}&=\int_{-\pi}^{\pi} \frac{dk}{2\pi}(\frac{\vert p(k)\vert^2}{\Delta-\omega(k)}+\frac{\vert q(k)\vert^2}{\Delta+\omega(k)})e^{ikx_{nm}}.
    \end{aligned}
    \end{equation}
When the giant atom only couple to single cell, the coupling coefficient can be written as $p(k)=\frac{1}{\sqrt{2}}(g_a+g_be^{i\phi(k)})$, $q(k)=\frac{1}{\sqrt{2}}(g_be^{i\phi(k)}-g_a)$.
Similar to Eq.(\ref{A1}), we can obtain the interaction in the topological phase 
     \begin{equation}
      J_{nm}=\left\{
        \begin{array}{rcl}
      &g_ag_b\frac{(-1)^{x_{nm}}}{J_1}(\frac{J_1}{J_2})^{x_{nm}}, \quad x_{nm}>0;\\
      &0, \quad\quad\quad\quad\quad\quad \quad\quad\quad\quad x_{nm}=0;\\
      &g_ag_b\frac{(-1)^{x_{nm}}}{J_1}(\frac{J_1}{J_2})^{-x_{nm}}, \quad x_{nm}<0;
    \end{array}%
      \right.
    \end{equation}%
In trivial phase,  we can obtain
    \begin{equation}
      J_{nm}=\left\{
        \begin{array}{rcl}
      &-g_ag_b\frac{(-1)^{x_{nm}}}{J_1}(\frac{J_2}{J_1})^{x_{nm}},\quad x_{nm}>0;\\
      &-\frac{2g_ag_b}{J1}, \quad\quad\quad \quad\quad\quad\quad x_{nm}=0;\\
      &-g_ag_b\frac{(-1)^{x_{nm}}}{J_1}(\frac{J_2}{J_1})^{-x_{nm}},  x_{nm}<0.
    \end{array}%
      \right.
    \end{equation}%

 \end{appendix} 
\bibliography{erery}
\end{document}